\documentclass[preprint,showpacs,preprintnumbers,amsmath,amssymb]{revtex4}

\usepackage{dcolumn}
\usepackage{bm}
\usepackage{graphicx}

\begin{document}

\title{Self-assembly of the simple cubic lattice with an isotropic potential}

\author{Mikael C. Rechtsman$^1$} \author{Frank H. Stillinger$^2$}
\author{Salvatore Torquato$^{2,3}$} 
\affiliation{$^1$Department of
Physics, Princeton University, Princeton, NJ 08544}
\affiliation{$^2$Department of Chemistry, Princeton University,
Princeton, NJ 08544} \affiliation{$^3$Program in Applied and
Computational Mathematics and PRISM, Princeton, NJ 08544}

\date{\today}

\begin{abstract}
Conventional wisdom presumes that low-coordinated crystal ground states require directional interactions.  Using our recently introduced optimization procedure to achieve self-assembly of targeted structures (Phys. Rev. Lett. 95, 228301 (2005), Phys. Rev. E 73, 011406 (2006)), we present an isotropic pair potential $V(r)$ for a three-dimensional many-particle system whose classical ground state is the low-coordinated simple cubic (SC) lattice.  This result is part of an ongoing pursuit by the authors to develop analytical and computational tools to solve statistical-mechanical inverse problems for the purpose of achieving targeted self-assembly.  The purpose of these methods is to design interparticle interactions that cause self-assembly of technologically important target structures for applications in photonics, catalysis, separation, sensors and electronics.  We also show that standard approximate integral-equation theories of the liquid state that utilize pair correlation function information cannot be used in the reverse mode to predict the correct simple cubic potential.  We report in passing optimized isotropic potentials that yield the body-centered cubic and simple hexagonal lattices, which provide other examples of non-close-packed structures that can be assembled using isotropic pair interactions.      
\end{abstract}

\maketitle

\section{Introduction}

There has been a tremendous amount of interest recently in nanoscale systems that assemble themselves into exotic and/or potentially technologically relevant structures.  The hope is that {\it self-assembly} will be the micro- and nano-fabrication process of the future, with devices being assembled more quickly and smaller than in the present microfabrication paradigm.  Self-assembly was originally defined (by Whitesides \cite{whitesides2}) as the process by which component parts (of any length scale) arrange themselves by virtue of their mutual, non-covalent interaction, into a larger functional unit.  Synthetic and biological examples abound, including colloidal crystallization and cluster formation \cite{Pine03}, DNA crystallization \cite{Winfree1}, organo-metallic patterning \cite{Andres1}, etc.  

By and large, the work by experimentalists and theorists on the problem of self-assembly has been based on trial and error, that is, there has been no systematic route developed to find the type of interparticle interaction needed to achieve given stuctures.  In previous work, we put forward computational tools based on inverse statistical mechanics \cite{RST-PRL1, RST-PRE1} that aim to solve exactly this problem.  In full generality, these methods find optimal interaction potentials among particles in a many-body system such that its classical ground state is a particular desired structure.  Specifically, we successfully applied them to two-dimensional, one-component systems wherein particles interact via isotropic pair potentials to produce the low-coordinated honeycomb and square lattices as our target structures.  

In this paper, we extend one of our algorithms to three dimensions in order to find an isotropic pair potential $V(r)$ for particles in an N-body system such that it will have the simple cubic lattice as its ground state.  It should be emphasized that low-coordinated crystals, such as the simple cubic lattice, have yet to be produced as ground states in three dimensions with isotropic pair potentials, to the best of our knowledge, due to the non-triviality of this task.  The simple cubic lattice's low density compared to the close packed structures, and its low coordination, mean that there is a real `balancing act' in choosing the correct functional form to cause this lattice to be the unique ground state.  This is most likely the reason that there has been no previously reported example of such a potential.

The present work connects closely to colloidal crystallization.  One well known technological application of the study of colloids is their capability to self-assemble into photonic crystals, which could be used in photonic devices.  Two examples of structures with favorable photonic properties are the diamond lattice \cite{photonics} and icosahedral quasicrystal \cite{ChaikinSteinhardt}.  Indeed, colloidal interactions can be tailored to some extent (through varying the particle charge, adjusting the solution salt concentration, dispersion and depletion interactions), with the goal of achieving a desired structure \cite{Russel}.  While extended colloidal quasicrystals seem very unlikely to be created in the lab, it is conceivable that colloidal particles can be designed to interact with each other isotropically in such a way that the diamond lattice would result given sufficient annealing time.  This study is a step towards a theoretical realization of that goal, as well as a non-trivial problem to test our optimization schemes.

In previous work \cite{RST-PRL1, RST-PRE1}, we discussed two numerical optimization schemes to find the desired potential.  Here, we employ one of these (the near-melting scheme) in order to find such an interaction.  The scheme will be discussed further in the following section.
The simple cubic lattice is 6-fold coordinated, and therefore this low coordinated structure is quite far away from the close-packed, 12-fold coordinated lattices such as the face-centered cubic (FCC) and hexagonally close-packed (HCP).  This suggests that the potential requires an attractive and repulsive component.  We include this information as input to the optimization scheme, with $V(r)$ being a member of a family of functions over which the optimization is carried out.  Our ultimate criterion for self-assembly is the very strong condition that perfect self-assembly be observed in a well-annealed molecular dynamics (MD) simulation.

Although the interaction potential for the simple cubic lattice (reported in the next section) was not intended for the atomic scale, it is interesting to note that metallic calcium (at high pressure) and polonium (low temperature, low pressure) both have simple cubic structures \cite{calcium1, polonium1}, and thus may have similar two-body pair potential approximants.  Although we make no attempt to compare the phase behavior of our model to that of calcium and polonium, this parallel suggests the possible applicability of simple isotropic potentials such as ours to model even atomic systems with open lattices.   

In the following section, we discuss the optimization of the potential and the results of the MD simulations.  We then discuss the use of liquid state properties for solving this inverse problem, specifically the use of integral equations.  Following that is a discussion and conclusion section in which the results are discussed in the context of self-assembly.  The appendix gives results obtained from the optimization schemes ($T=0$) for two different lattices, namely the body-centered cubic (BCC) and simple hexagonal lattices.  We note that these potentials are significantly and qualitatively different from the simple cubic lattice potential.

\section{Optimization Procedure}

In order to find a spherically symmetric pair interaction potential that yields the simple cubic lattice as its ground state, we use an optimization scheme developed by the authors \cite{RST-PRL1, RST-PRE1}, straightforwardly generalized to three dimensions.  In particular, we employ the `near-melting' scheme that is based on minimizing the Lindemann parameter of the crystal, just below its melting point, within a family of potentials $V(r;\{a_0...a_n\})$ described by the parameters $\{a_0...a_n\}$.  The optimized potential will be called $V_{SC}(r)$. After this optimized solution is produced by the program, we check three necessary conditions that the potential must satisfy, namely that: (1) it is energetically favored over its competitors for a non-trivial specific volume range, (2) there exists a range of stability in pressure at zero temperature (from the Maxwell double-tangent construction), and (3) all phonon frequencies are real.  If any of these conditions are not met, it is impossible for the simple cubic crystal to be a ground state.  Taken together, these conditions are strongly suggestive that a working solution has been found to this inverse problem.  Note here that the competitor lattices considered are the face-centered cubic, hexagonally-close-packed, body-centered cubic, simple hexagonal (with axial ratio, $c/a$, 1), diamond, and wurtzite \cite{WurtziteRef} crystals. 

Provided the aforementioned conditions are met, we run molecular dynamics (MD) simulations of the particles interacting via the optimized potential, $V_{SC}(r)$, in the microcanonical (number-volume-energy, or NVE) ensemble.  The standard Verlet algorithm is used.  The system, typically composed of $\sim 200$ particles, is slowly cooled by velocity rescaling through its freezing transition and slowly annealed until as many defects are removed as is practical in reasonable computer time.  We employ periodic boundary conditions in a cubic simulation box.  If the simple cubic structure is produced, we take this as the decisive piece of evidence that this structure is indeed the ground state of $V_{SC}$.  For the remainder of this section, we will discuss outcome of the optimization scheme, the properties of the potential, and the results of the MD simulation.

As input to the optimization, the program takes a parameterized function $V(r;\{a_0...a_n\})$.  The simple cubic lattice has 6 nearest neighbors, 12 second neighbors and 8 third neighbors at distances unity, $\sqrt 2$ and $\sqrt 3$, respectively.  Motivated by this, we choose a parameterization that allows for the first neighbor to be energetically suppressed (i.e. $V(1)>0$), and the second and third neighbors to be either favored or suppressed.  The parameterization we chose was
\begin{equation}
V(r;\{a_0...a_3\}) = \frac{1}{r^{12}} - 50\frac{a_0}{a_1}\exp[{-a_1(r-\sqrt 2)^2}] - 25\frac{a_2}{a_3}\exp[{-a_3(r-\sqrt 3)^2}]. \label{parameterization}
\end{equation} 
A number of MD simulations were run on members of this family of potentials and it was found that for some, structures bearing close resemblance to the simple cubic would emerge, albeit with stacking faults and other defects present.  This was taken as motivation to proceed with this family.  In order to remove the effect of `over-stiffening' in the optimization procedure (in which a region in parameter space is found in which the crystal is very stiff and thus away from melting), we put two constraints on the parameters, which reduces the number of degrees of freedom in the optimization from 4 to 2.  These constraints are
\begin{eqnarray}
a_0 + a_2 = 2\\
a_1 + a_3 = 100
\end{eqnarray}
These constraints are introduced to prevent an unbounded rescaling of the two Gaussian functions, as well as to essentially restrict either Gaussian from becoming arbitrarily narrow, thus creating an undesirably high crystal stiffness. Due to the small number of degrees of freedom, a global search of parameter space (within reasonable bounds) was performed.  The parameters output were $a_0=1.7762$, $a_1=32.2844$, $a_2 = 0.2238$, $a_3=67.7156$.  This gives a result for the simple cubic potential.  Upon examining this function, it can be seen that the second Gaussian (the one centered at $\sqrt 3$) produces a qualitatively small effect on the shape of the function.  Therefore, in the interest of finding the simplest possible potential to stabilize the simple cubic structure, we ran MD simulations using this potential with and without the second Gaussian included.  In our simulations, that Gaussian seemed to have no effect whatsoever on the final result, and so we choose to drop it entirely.  Thus, the final result for the potential is 
\begin{equation}
V_{SC}(r) =  \frac{1}{r^{12}} - 2.7509\exp[-32.2844(r-\sqrt 2)^2].\label{sc-eqn}
\end{equation}    

\begin{figure}
\includegraphics[scale=0.35,clip,viewport=0pt 0pt 717pt 535pt]{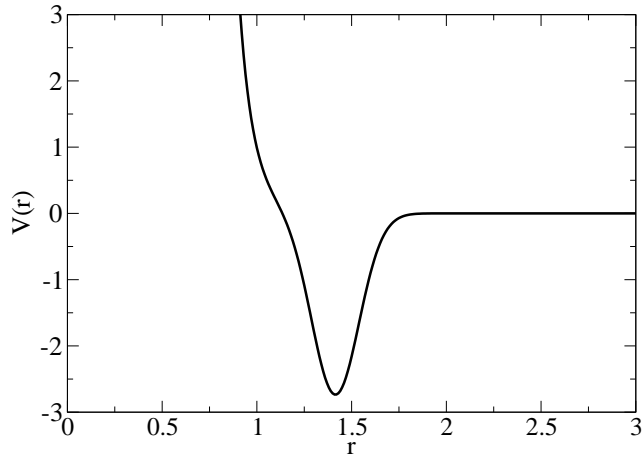}
\caption{\label{fig:potential} $V_{SC}(r)$, given in Eq. (\ref{sc-eqn}), the isotropic interaction potential that yields the simple cubic lattice as the ground state. }
\end{figure}

\begin{figure}
\includegraphics[scale=0.35,clip,viewport=0pt 0pt 717pt 535pt]{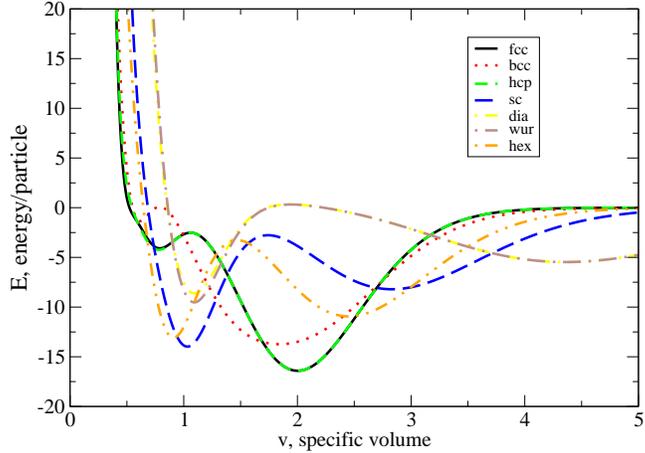}
\caption{\label{fig:latticesums} (Color online) Lattice sums for the simple cubic potential, $V_{SC}(r)$, given in Eq. (\ref{sc-eqn}).  The simple cubic lattice has a pressure range of stability of $p=2.64$ through $p=7.89$, and a specific volume range of stability of $v=0.98$ through $v=1.03$.}
\end{figure}

\begin{figure}
\includegraphics[scale=0.35,clip,viewport=0pt 0pt 717pt 535pt]{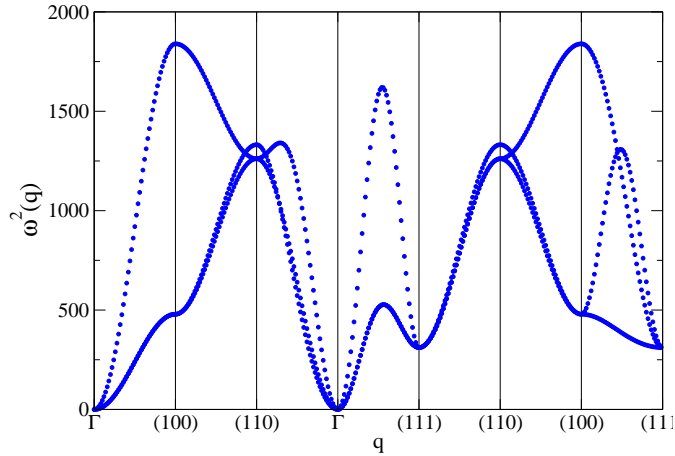}
\caption{\label{fig:phononspectrum} Phonon spectrum for the simple cubic lattice interacting via $V_{SC}(r)$, given in Eq. (\ref{sc-eqn}), at specific volume $v=1.0$.  Points of high symmetry in the Brillouin zone are indicated by vertical lines in the plot; they are given by their Miller indices.  A straight line through reciprocal space connects each high symmetry point. }
\end{figure}

This potential is shown in Fig. \ref{fig:potential}.  The form of the potential is such that the first neighbor, which is 6-fold coordinated, lies at a distance at which the pair potential is positive.  This inhibits the formation of the 12-coordinated close-packed lattices.  The deep negative minimum (the gaussian) is centered at the second neighbor distance, $\sqrt 2$, which is 12-fold coordinated and thus strongly favored.   To explicitly show that the necessary conditions for self-assembly of the simple cubic lattice are met, we show the lattice sums in Fig. \ref{fig:latticesums} and the phonon spectrum in Fig. \ref{fig:phononspectrum}.  In the plot of the lattice sums, it is seen that the simple cubic is energetically favored for a non-trivial range of specific volume, as compared to the competitor lattices previously enumerated.  Specifically, we can determine from the double-tangent construction that at $T=0$, the simple cubic lattice has a pressure range of stability of $p=2.64$ through $p=7.89$, and a specific volume range of stability of $v=0.98$ through $v=1.03$.  The pressure, for a given lattice, is the negative slope of its energy curve in the lattice sums, namely $-d\epsilon/dv$ where $\epsilon$ is the energy per particle and $v$ is the specific volume.   This is at zero temperature, at which the lattice sums are performed.  Length and energy units are defined by the axes of Fig. \ref{fig:potential}.  In the plot of the phonon spectrum, which is at specific volume unity, we plot the square of the crystal frequency $\omega^2({\bf q})$, for a selection of points in the Brillouin zone of the simple cubic lattice.           

The potential given in Eq. (\ref{sc-eqn}) is by no means unique; it is merely the optimal form within the family of functions defined by Eq. (\ref{parameterization}), according to our criteria for self-assembly.  Of course there is an infinite number of potentials that would correspond to a particular ground state structure, each of which have different finite-temperature structural behavior.  That said, the ground state structure is a key property for experimentalists, as in many colloidal systems, the characteristic strength of the interaction potential is much larger than $k_BT$, so the ground state is highly relevant.

\begin{figure}
\includegraphics[scale=0.35,clip,viewport=0pt 0pt 717pt 535pt]{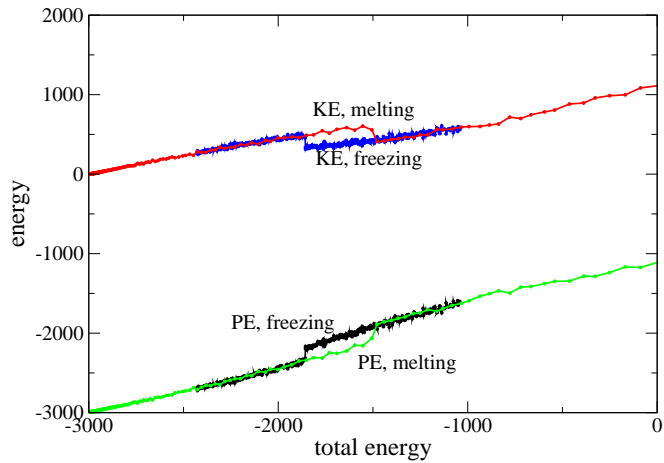}
\caption{\label{fig:hysteresis} (Color online) Kinetic energy (KE) and potential energy (PE) plotted against total energy in two different MD simulations (freezing of the liquid and melting of the crystal), employing the potential of Eq. \ref{sc-eqn}.  The liquid supercools but then undergoes complete nucleation of the simple cubic lattice.  }
\end{figure}

\begin{figure}
\includegraphics[scale=0.35,clip,viewport=0pt 0pt 717pt 535pt]{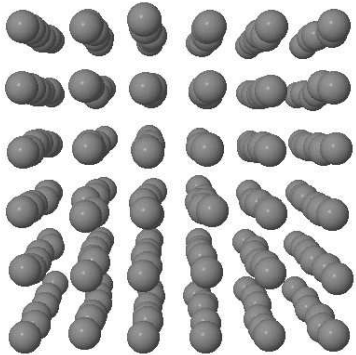}
\caption{\label{fig:md} Snapshot of the simple cubic structure produced upon slow cooling with particles interacting via $V_{SC}$.  The system is shown at a small positive temperature, thus small harmonic fluctuations about the equilibrium positions are present.  The simple cubic lattice has a pressure range of stability of $p=2.64$ through $p=7.89$, and a specific volume range of stability of $v=0.98$ through $v=1.03$.}
\end{figure}

During the MD simulation in which the liquid freezes into the simple cubic crystal, we track potential and kinetic energies, as shown in Fig. \ref{fig:hysteresis}.  Also in this figure is a MD simulation of the perfect crystal being heated and melting into the liquid.  It is clear that there is significant hysteresis present; we believe the liquid supercools, rather than the solid overheating.  Figure \ref{fig:md} shows the simple cubic lattice produced upon slow cooling from a random initial configuration (216 particles, specific volume $v=1.0$).  The geometry of the box is such that $6^3=216$ particles is a `magic number' for the simple cubic lattice, allowing it to fit naturally into the box.  This implies that if the box is replicated throughout space, a perfect simple cubic lattice results.  At a constant pressure of $p=3.5$, within the pressure range of stability given by the double tangent construction applied to the $T=0$ lattice sums, we find that upon melting the volume expands by $88.$ units, a factor of approximately $4/3$.  As mentioned, all units are defined in terms of Fig. \ref{fig:potential}, which sets the length and energy scales.  This calculation was carried out in an $NPT$-ensemble Monte Carlo simulation.

\section{Remarks about the Liquid State}

In order to study the liquid state of a system interacting via $V_{SC}$, we have run a NVT canonical Monte Carlo simulation of such a system just above its freezing point, at $k_BT=2.0$.  From this, we extracted the radial distribution function, $g(r)$, of the liquid.  We applied the hypernetted-chain (HNC) and Percus-Yevick (PY) approximations for $g(r)$ as input to find corresponding approximations for the pair potentials, $V_{HNC}$ and $V_{PY}$, respectively. The starting point of this analysis is the Ornstein-Zernike integral equation,

\begin{equation}
h({\bf r}) = c({\bf r}) + \int d{\bf r}^\prime c({\bf r}^\prime)h(|{\bf r}-{\bf r}^\prime|)\label{oz}
\end{equation}
where $h({\bf r})=g({\bf r})-1$, and $c({\bf r})$ is the direct correlation function, defined by relation \ref{oz}.  The HNC and PY approximations are closures to this equation, and are given by, respectively,
\begin{eqnarray}
c_{HNC}({\bf r}) = g({\bf r})-1-\ln g({\bf r}) - V_{HNC}({\bf r})/k_BT\\
c_{PY}({\bf r}) = g({\bf r})-\left[g({\bf r})\exp (V_{PY}({\bf r})/k_BT)\right]
\end{eqnarray}    

\begin{figure}
\includegraphics[scale=0.35,clip,viewport=0pt 0pt 717pt 535pt]{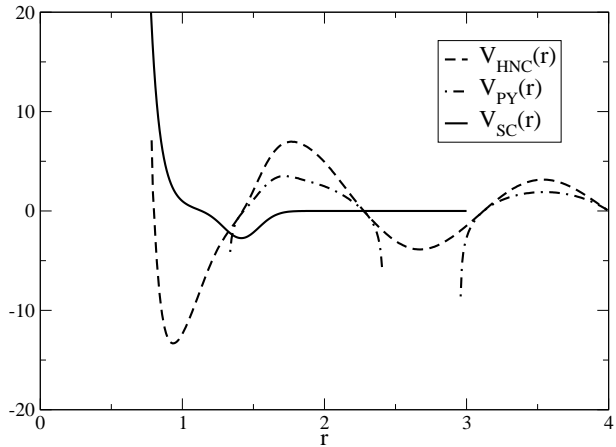}
\caption{\label{fig:liquidstate} Interaction potentials obtained from Ornstein-Zernike integral equation via HNC and PY closures, compared to the actual simple cubic potential \ref{sc-eqn}.  Clearly they differ greatly from the actual interaction potential, $V_{SC}(r)$.}
\end{figure}

The result of this analysis is shown in Fig. \ref{fig:liquidstate}.  As can be seen in Fig. \ref{fig:liquidstate}, there are points at which the HNC and PY approximations do not exist because at these points the approximation calls for the natural logarithm of a negative number.  The HNC solution appears to have long-ranged oscillations that seem to be irrelevant to an SC potential.  A fit to the HNC potential was obtained, and lattice sums were carried out on the simple cubic and all the aforementioned competitor lattices, given previously.  Since the $V_{HNC}$ is only defined for $r>0.76$, we only considered densities for which each lattice had no neighbors at those distances.   According to these lattice sums, the simple cubic lattice was not stable at any value of the specific volume.  At $v=1.0$, the specific volume at which the integral equations were solved, the FCC was the lattice of lowest energy.  Although there were two specific volume ranges for which the simple cubic corresponded to the lowest energy, the double-tangent criterion was not satisfied there.  Clearly the liquid state approximations produce potentials entirely different from the true $V_{SC}$.  Although use of the HNC and PY closures in the conventional manner to predict pair correlation in the liquid for a given pair potential is demonstrably useful, inverting the procedure as we have attempted here to infer an appropriate $V(r)$ for a targeted crystal is not fruitful.

\section{Conclusions and Discussion}

In summary, we have optimized for an isotropic interaction potential that has the simple cubic lattice as its ground state.  A few words in regards to the optimization schemes are called for here.  As mentioned briefly earlier in this work (and discussed in detail in Ref. \cite{RST-PRE1}), we employ two optimization schemes to find suitable interaction potentials.  These two are the `zero-temperature' scheme, in which lattice energy of the desired target lattice is minimized subject to the constraint that the lattice is linearly stable (real phonon frequencies), and the `near-melting' scheme, in which the Lindemann parameter of the potential is minimized in order to take into account anharmonic behavior (vacancies, interstitials, liquid nucleation).  The prescribed method is to first apply the zero-temperature scheme in order to find a potential that has the target lattice energetically favored and is mechanically stable, and only then to apply the near-melting scheme over a relatively small region of parameter space.  Seldom are both schemes needed, however.  For example, when the first scheme produces a potential that itself causes assembly of the target structure, then the second is unnecessary (as in the case of the BCC and simple hexagonal), or when a potential can be guessed such that the lattice sums are already favorable and the target lattice is mechanically stable, as in the case of $V_{SC}$.

At the heart of our study is the question of exploring the limitations of isotropic interactions.  Specifically, we would like to know how complex ground state structures can be with only isotropic interaction.  The difficulty of finding a potential for a given structure may be cast in terms of its specific volume, or perhaps coordination number: close-packed lattices, such as FCC and HCP, which are 12-fold coordinated and have specific volume $0.7071n^3$ ($n$ is the nearest neighbor distance), can be assembled by the oft-cited Lennard-Jones potential (HCP), among many others (including the gaussian-core model at low density (FCC) \cite{gaussiancore} and the simple $1/r^{12}$ potential).  The BCC, 8-fold coordinated and with specific volume $0.7698n^3$, is accepted to be the ground state structure of the gaussian core model at high densities, and the Coulomb potential, for example.  The simple hexagonal lattice is also 8-fold coordinated, with specific volume $0.8660n^3$ (less than BCC), and seems in the authors experience to be more sensitive (in terms of its self-assembly) to changes in the functional form of the potential.  The simple cubic lattice is 6-fold coordinated and has specific volume $v=1.000n^3$, and has required a very careful choice of parameterization of the family of potentials to find a $V(r)$ that yields self-assembly of that lattice.  The difficulty of finding a suitable potential may indeed come from the larger number of intervening structures as the specific volume of the target lattice is increased.  We should stress here that this is not a strict rule, but only to provide insight on the significant non-triviality of the problem of assembling open structures with isotropic potentials.   In this sense, the diamond should be the most challenging lattice, and is the next natural step in our pursuit of exotic ground states of isotropic potentials.  As stated earlier, diamond lattice assembly could have technological applications since dielectric colloidal spheres in such an arrangement would have a photonic bandgap across the Brillouin zone \cite{photonics}, and would thus be a photonic crystal.  In future work, we also plan on searching for more exotic ground state structures, for example helices, three-dimensional arrays of chains, and multiscale structures.

\section{Appendix: Body-Centered and Simple Hexagonal Lattices}

\begin{figure}
\includegraphics[scale=0.35,clip,viewport=0pt 0pt 717pt 535pt]{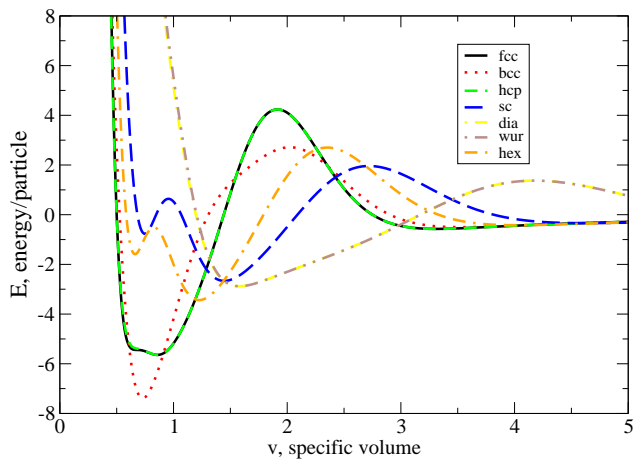}
\caption{\label{fig:BCC-ls} (Color online) Lattice sums for the BCC potential, $V_{BCC}$, given in Eq. (\ref{bcc-equ}).  The BCC is the global ground state structure (for any value of specific volume).}
\end{figure}

\begin{figure}
\includegraphics[scale=0.35,clip,viewport=0pt 0pt 717pt 535pt]{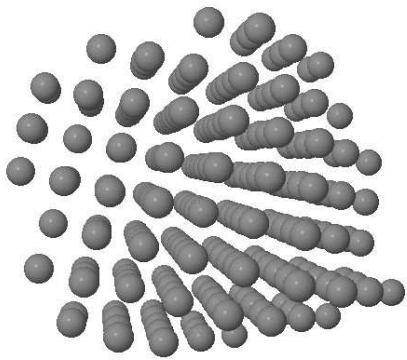}
\caption{\label{fig:BCC-md} Snapshot of a 216-particle MD simulation of particles interacting via the $V_{BCC}$ potential at specific volume $v=0.729$.  This is a perfect BCC lattice, produced on slow cooling of the system.  The pressure range of stability of the BCC lattice is $0.0$ through $24.3$, and the specific volume range is $0.65$ through $0.73$, at zero temperature.}
\end{figure}

\begin{figure}
\includegraphics[scale=0.35,clip,viewport=0pt 0pt 717pt 535pt]{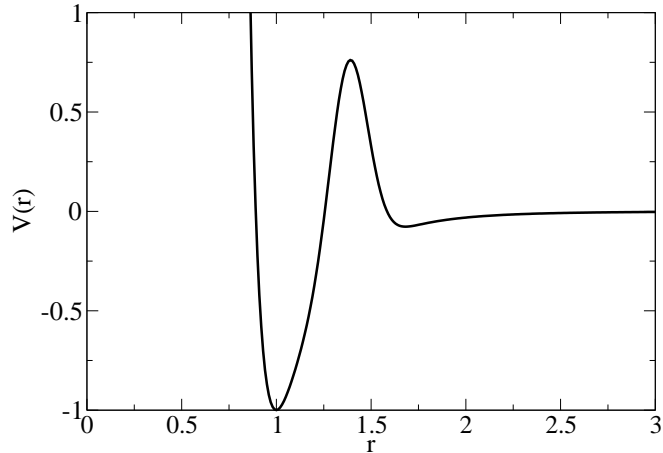}
\caption{\label{fig:BCC-pot} $V_{BCC}(r)$, given in Eq. (\ref{bcc-equ}), the isotropic interaction potential that yields the body-centered cubic lattice as the ground state. }
\end{figure}

In this section we report potentials, obtained by inverse optimization, that cause spontaneous self-assembly of the body-centered cubic (BCC) and simple hexagonal lattices.  The simple hexagonal lattice is composed of layers of triangular lattice stacked directly on top of one another, with interlayer distance equal to the in-plane nearest neighbor distance.  Here we apply the `zero-temperature' optimization procedure, which will not be expounded upon here since it was reported in detail in Ref. \cite{RST-PRE1}, and described briefly in the previous section.
The pair potential for the BCC lattice, $V_{BCC}(r)$, is given by
\begin{equation}
V_{BCC}(r) = \frac{1}{r^{12}}-\frac{2}{r^6} + 1.023\exp\left[-52.0(r-1.382)^2\right] \label{bcc-equ}
\end{equation}
and is shown in Fig. \ref{fig:BCC-pot}.  The lattice sums for the BCC lattice are shown in Fig. \ref{fig:BCC-ls}, and the result of a slowly cooled NVE MD run is shown in Fig \ref{fig:BCC-md}.  This is a $216$ particle configuration in which a defect-free BCC lattice emerges upon slow cooling and annealing.  The pressure range of stability of the BCC lattice is $0.0$ through $24.3$, and the specific volume range is $0.65$ through $0.73$, at zero temperature, according to the lattice sums.   

\begin{figure}
\includegraphics[scale=0.35,clip,viewport=0pt 0pt 717pt 535pt]{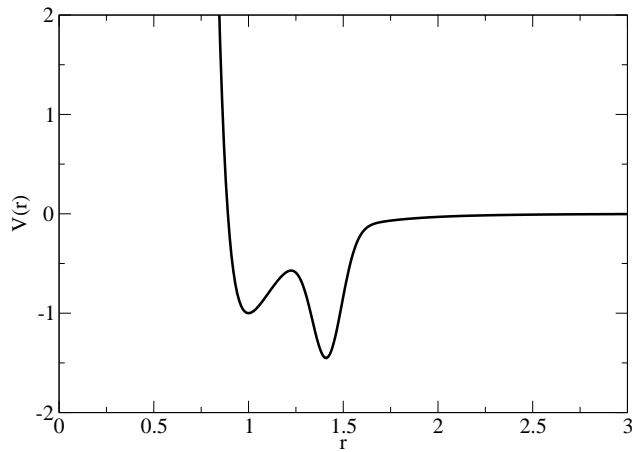}
\caption{\label{fig:SHEX-pot} $V_{SHEX}(r)$, given in Eq. (\ref{shex-equ}), the isotropic interaction potential that yields the simple hexagonal lattice as the ground state. }
\end{figure}

\begin{figure}
\includegraphics[scale=0.35,clip,viewport=0pt 0pt 717pt 535pt]{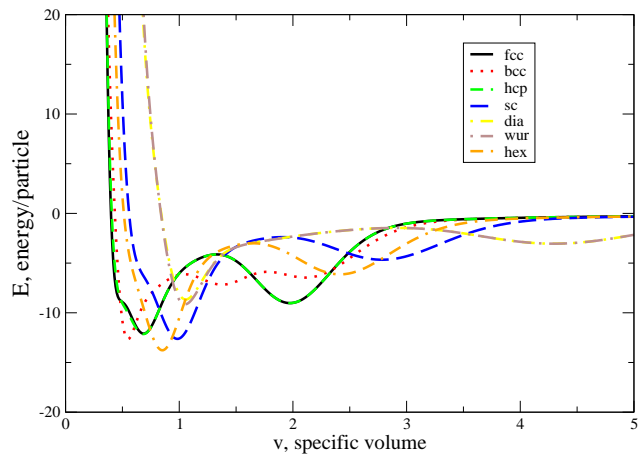}
\caption{\label{fig:SHEX-ls} (Color online) Lattice sums for the SHEX potential, $V_{SHEX}(r)$, given in Eq. (\ref{shex-equ}).  The SHEX is the global ground state structure (for any value of specific volume).}
\end{figure}

\begin{figure}
\includegraphics[scale=0.35,clip,viewport=0pt 0pt 717pt 535pt]{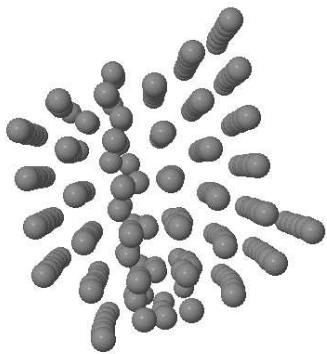}
\caption{\label{fig:SHEX-mc} Snapshot of a 216-particle MC simulation of particles interacting via the $V_{SHEX}$ potential at specific volume $v=0.866$.  This is a predominantly simple hexagonal configuration, with a single stacking fault that is present as a result of the misalignment of particles within the periodic box.  This structure was produced upon slow cooling of the system.  The pressure range of stability is $0.0$ through $6.1$, and the specific volume range is $0.82$ through $0.86$.}
\end{figure}

The pair potential for the simple hexagonal lattice, $V_{SHEX}(r)$, is given by
\begin{equation}
V_{SHEX}(r) = \frac{1}{r^{12}}-\frac{2}{r^6} -1.215\exp\left[-81.37(r-\sqrt{2})^2\right] \label{shex-equ}
\end{equation}
and is shown in Fig. \ref{fig:SHEX-pot}. The lattice sums for the simple hexagonal lattice are shown in Fig. \ref{fig:SHEX-ls}, and the result of a slowly cooled NVT Monte Carlo (MC) run is shown in Fig \ref{fig:SHEX-mc}.  This is a $216$ particle configuration in which a defect-free simple hexagonal lattice emerges upon slow cooling and annealing.  The pressure range of stability is $0.0$ through $6.1$, and the specific volume range is $0.82$ through $0.86$.

Although pair potentials have been previously discovered (although not through optimization) that yield these lattices as ground states \cite{stillinger:5095,laviolette:3335}, these cases provide non-trivial test cases for our optimization schemes.  It should be noted that in Ref. \cite{laviolette:3335}, the authors find a hexagonal structure whose c-to-a ratio is not exactly unity; thus it is not the perfect simple hexagonal lattice.  Furthermore, the optimization schemes provide a means for obtaining the potential that most robustly stabilizes the desired target lattice, however that is to be defined (e.g. stable against density perturbations, or those in pressure, or the functional form of $V(r)$).

\end{document}